\documentclass[nato,numreferences]{crckbked}

\usepackage{graphicx}
\numreferences
\usepackage{klucite}

\makeindex

\begin{document}

\begin{article}

\begin{opening}
\title{Quantum fluctuations and electronic transport through strongly
  interacting quantum dots}

\author{T. A. \surname{Costi}\email{tac@tkm.physik.uni-karlsruhe.de}}
\institute{Universit\"at Karlsruhe, Institut f\"ur Theorie der Kondensierten
  Materie 76128 Karlsruhe, Germany}

\runningauthor{T. A. Costi}

\begin{abstract}
We study electronic transport through a strongly
interacting quantum dot by using the finite temperature extension of Wilson's
numerical renormalization group (NRG) method. This allows the linear 
conductance to be calculated at all temperatures and in particular 
at very low temperature where quantum fluctuations and the Kondo effect
strongly modify the transport. The quantum dot investigated has 
one active level for transport and is modeled by an Anderson impurity model
attached to left and right electron reservoirs. The predictions for the linear
conductance are compared to available experimental data for quantum dots in
heterostructures. The spin-resolved conductance is calculated as
a function of gate voltage, temperature and magnetic field strength and the
spin-filtering properties of quantum dots in a magnetic field are
described.
\end{abstract}
\end{opening}

\section{Introduction}

Recent experimental work 
\cite{goldhaber.98,cronenwett.98,schmid.98,simmel.99,wiel.00}  
has demonstrated the importance of correlations in determining the low 
temperature transport properties of nanoscale size quantum dots.
These dots consists of a confined region of electrons, of typical
diameter $100nm$, ``weakly'' coupled to leads via tunnel barriers. 
A ``weak'' coupling, $\Gamma$, means that the quantized levels of the
dot are broadened into resonances, but are not completely washed out. 
A gate voltage, $V_g$, controls the position of the quantized levels 
relative to the chemical potentials
of the leads and thereby the total number of electrons on the dot. 
The charging energy $U$ for adding electrons to the dot from the surrounding 
electron reservoirs (leads) is typically the largest energy scale, and the
dot is strongly correlated provided $U/\Gamma \gg 1$. Typical values for $U$ 
and $\Gamma$ for the dots in \cite{goldhaber.98,wiel.00} are $U\sim 0.5-2.0$ meV 
and $\Gamma \sim 0.2-0.3$ meV, so these dots are strongly correlated.

For temperatures $\Gamma \ll T\ll U$ quantum fluctuations are small, and
transport is dominated by charging effects. This regime is well understood
\cite{beenakker.91}. The conductance, $G$, exhibits a series of 
approximately equidistant peaks as a function of $V_g$, with spacing $U$.
The peaks correspond to a fractional number of electrons on the dot and
alternate ``Coulomb blockade'' valleys to either an even or an 
odd number of electrons.

In this paper we shall be interested in the regime $T\lsim \Gamma \ll U$, where 
the strong quantum fluctuations can lead to a dramatic modification 
of the above picture of Coulomb blockade. 
In particular, for an odd number of electrons on the dot, 
the quantum dot can have a net spin $1/2$, and a Kondo effect can develop. 
It has been predicted \cite{glazman.88,ng.88}, that, at low temperature, 
this enhances the conductance in the odd electron valleys, 
turning them instead into plateaus of near perfect transmission. 

The outline of the paper is as follows. Sec.~\ref{model} describes the model
and Sec.~\ref{nrg-method} 
the NRG method used to solve it. The results \cite{costi.94,costi.00,costi.01}
for the conductance are described in Sec.~\ref{zero-field}-\ref{scaling} 
and these are used to compare with experiment over a 
range of gate voltages and temperatures in Sec.~\ref{comparison}.  
Experimental investigation 
of the regime $T\lsim \Gamma$ has only recently become possible as a result of 
better control of the electrode geometry, allowing parameters like $\Gamma$ to 
be tuned to values of $1$-$3$K so that $T\lsim\Gamma$ is accessible
\cite{goldhaber.98,cronenwett.98,schmid.98,simmel.99,wiel.00}. 
Sec.~\ref{magnetic-field-effects} 
describes the effect of a magnetic field on the transport through 
a quantum dot. Influencing transport via spin effects is of great current
interest \cite{ohno.99}. The results presented in Sec.~\ref{magnetic-field-effects} 
should prove useful for interpreting magnetotransport experiments on 
strongly interacting quantum dots.

\section{Model}
\label{model}
At sufficiently low temperature, transport through a strongly interacting
quantum dot will be mainly determined by the partially occupied level of the 
dot, denoted $\varepsilon_d$, which lies closest to the chemical potential 
of the leads. Its occupancy, $n_d$, can be varied from $n_d=0$
to $n_d=2$ by varying the gate voltage $V_g$, where $-eV_g=\varepsilon_d$.
The resulting model of a single correlated level $\varepsilon_{d}$ 
with Coulomb repulsion $U$, coupled to left and right free electron 
reservoirs can be written as
\begin{eqnarray}
{\cal H} &=& \sum_{\sigma}\varepsilon_{d}d_{\sigma}^{\dagger}d_{\sigma} 
+ Ud_{\uparrow}^{\dagger}d_{\uparrow}d_{\downarrow}^{\dagger}d_{\downarrow}
+ g\mu_{B}HS_{z}^{d}\nonumber\\
  &+& \sum_{k,\sigma,i=L,R}\varepsilon_{k,i}c_{k,i,\sigma}^{\dagger}c_{k,i,\sigma}+
         \sum_{k,\sigma,i=L,R}V_{i}(c_{k,i,\sigma}^{\dagger}d_{\sigma} +
          d_{\sigma}^{\dagger}c_{k,i,\sigma}).\label{eq:AM}
\end{eqnarray}
We assume energy independent lead couplings $\Gamma_{L,R}=2\pi\rho_{L,R}(\epsilon_F) 
V_{L,R}^2$, where $\rho_{L,R}(\epsilon_F)$ is the Fermi level 
density of states (per spin) of the left/right electron reservoir.
The first two terms in ${\cal H}$ represent the quantum dot, the third term 
is a magnetic field coupling only to the dot's spin
$S_{z}^{d}=\frac{1}{2}(d_{\uparrow}^{\dagger}d_{\uparrow}
-d_{\downarrow}^{\dagger}d_{\downarrow})$ (we set $g=\mu_B = 1$), the fourth term 
represents the free electron reservoirs and the last
term is the coupling between the dot and the reservoirs.
This model can be reduced to the standard Anderson impurity model of a single reservoir
attached to the dot with strength $\Gamma=\Gamma_{L}+\Gamma_{R}$
\cite{glazman.88}. Note that $\Gamma=2\Delta$, where $\Delta$ is the
hybridization strength as usually defined in the Anderson model
\cite{hewson.93}. We use $\Gamma$ throughout. In experiments 
\cite{goldhaber.98, wiel.00}, $\Gamma$ is extracted by analyzing the 
high temperature ($T\gg \Gamma$) behaviour of the conductance peaks (see
figure:~\ref{figure2}a below).

We assume, from here on, symmetric coupling 
to the leads, $\Gamma_L=\Gamma_R$. The linear magnetoconductance, 
$G(T,H)=\sum_{\sigma}G_{\sigma}(T,H)$, is written
as a sum of spin-resolved magnetoconductances, $G_{\sigma}$, where
\begin{eqnarray}
G_{\sigma}(T,H) &=&\frac{e^2}{\hbar}\Gamma
\int_{-\infty}^{+\infty} d\omega\; A_{\sigma}(\omega,T,H)
\left(-\frac{\partial f(\omega)}{\partial\omega}\right).\label{eq:conductance}
\end{eqnarray}
$A_{\sigma}(\omega,T,H)$ 
is the equilibrium spectral density and is expressed in terms of the 
local level Green function, ${\cal G}_{d,\sigma}=1/(\omega-\varepsilon_d +i\Gamma/2 -
\Sigma_U)$, with $\Sigma_U$ the correlation part of the self-energy, by
\begin{eqnarray}
&&A_{\sigma}(\omega,T,H)
=-\frac{1}{\pi}{\rm Im}\;
{\cal G}_{d,\sigma}(\omega+i\epsilon,T,H),
\label{eq:spin-resolved-sd}
\end{eqnarray}

\section{Method}
\label{nrg-method}
We calculate $A_{\sigma}(\omega,T,H)$ by using
Wilson's NRG method \cite{wilson.75+kww.80} extended
to finite temperature dynamics\cite{costi.94,costi.99},
with recent refinements \cite{bulla.98,hofstetter.00} which
improve the high energy features. The NRG procedure for finite
temperature dynamics is described in \cite{costi.94,costi.99}. Here, we
make a few remarks concerning the above refinements.

The first refinement \cite{bulla.98} uses the correlation part of the 
self-energy, $\Sigma_U$ , to evaluate the spectral density $A$ in the Anderson
impurity model \cite{bulla.98}. This improves the spectra around the single-particle
excitations $\varepsilon_d$ and $\varepsilon_d + U$ since the single particle
broadening $\Gamma/2$ 
is put into the Green function, ${\cal G}_d$, exactly, thereby making the
excitations at $\varepsilon_d$ and $\varepsilon_d + U$ slightly sharper than
in the earlier procedure \cite{costi.94} which evaluated $A$ directly from its
spectral representation. The description of the low energy Kondo
resonance is equally accurate within both approaches and close to 
exact\cite{bulla.98}.

The second refinement uses the reduced density matrix in the NRG procedure
for dynamical quantities \cite{hofstetter.00} in place 
of the grand canonical density matrix of the usual procedure \cite{costi.94}. 
This reduces finite-size effects in the spectra. The latter are usually
small on all energy scales in the case of zero magnetic field, but can be large
in the case of a finite magnetic field, particularly for the high energy parts
of the spectra (since these are calculated from the shortest chains and are 
therefore subject to the largest finite-size effects).
\begin{figure}[h]
\centerline{\includegraphics[width=6cm]{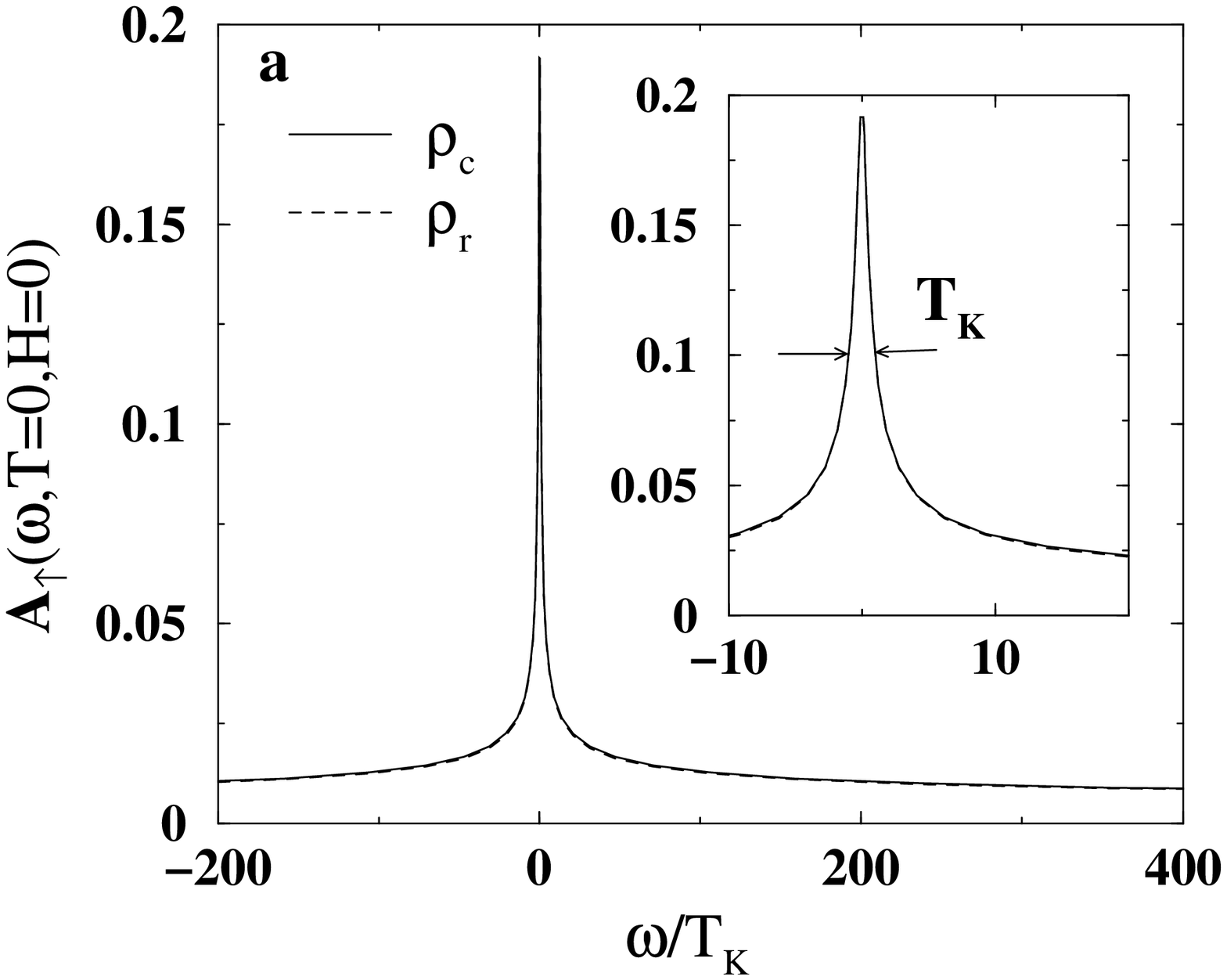}
\includegraphics[width=6cm]{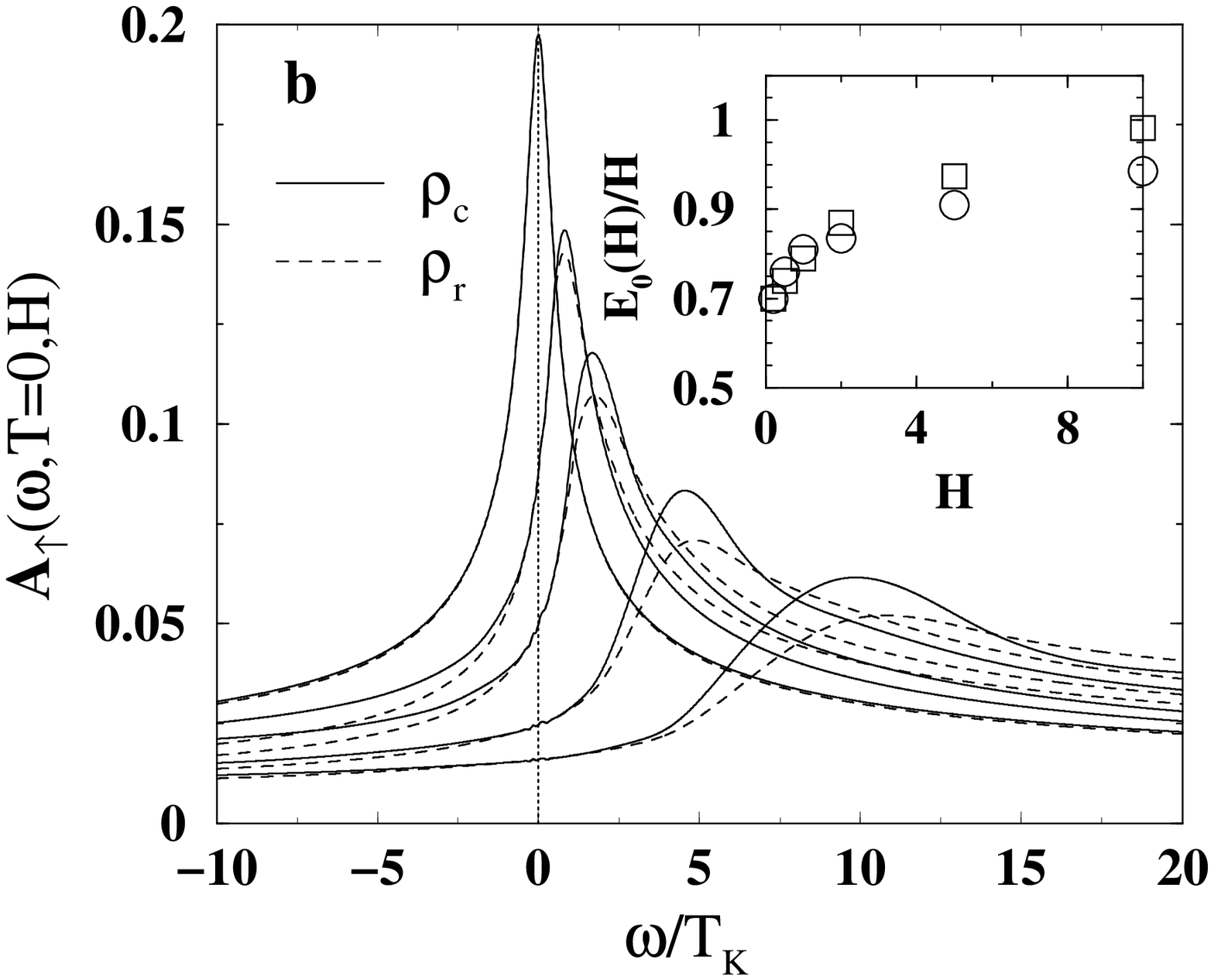}}
\caption{Comparison of methods for $A_{\uparrow}(\omega,0,H)$,
of the Kondo model, using canonical ($\rho_c$, solid line ) 
    \cite{costi.94} and reduced density matrices ($\rho_r$, dashed line)
  \cite{hofstetter.00}.
  $T_K=3.97\times 10^{-5}$ (in units of half bandwidth $D=1$) is the
  HWHM of the $T=H=0$ Kondo resonance : (a) at zero field, with the 
  inset showing the Kondo resonance in more detail, and, 
  (b) in finite field, with the curves 
  from left to right corresponding to $H= 0, 1, 2, 5, 10$ in units of $T_K$.
  The inset shows the peak position, $E_0(H)$, 
  as a function of $H$ for the canonical (circles) and reduced density matrix
  (squares) approaches. Both $E_0$ and $H$ are measured in units of $T_K$. The
  limit \cite{logan.01} $\lim_{H\rightarrow 0}E_0(H)/H = 2/3$ is 
  recovered in both approaches.
  }
\label{figure1}
\end{figure}
Figure:~\ref{figure1}a shows that both procedures give equally accurate results 
for the Kondo model in zero magnetic field on all energy scales.
The same holds for the Anderson model in zero magnetic field. At finite
magnetic field, the reduced density matrix is required for a correct description
of the features at $\varepsilon_d$ and $\varepsilon_d + U$ in the spin-resolved
spectra \cite{hofstetter.00}. In contrast, the low energy Kondo resonance in a 
magnetic field is well described by either procedure for $H\lsim 10T_K$, as shown in 
figure:~\ref{figure1}b for the Kondo model.

The position, $E_0(H)$, of the spin-resolved Kondo resonance in a weak 
magnetic field, $H\ll T_K$, is known from an exact Fermi liquid 
result due to Logan \cite{logan.01} (see also \cite{hewson.01} showing the
low field quasiparticle resonance).
This states that $\lim_{H\rightarrow 0}E_0(H)/H = 2/3$. This indicates
that correlations reduce $E_{0}(H)/H$ from its expected high field 
limit of $1$ to $2/3$. The inset to  figure:~\ref{figure1}b shows that
both procedures recover this result. 
The use of the reduced density matrix has also proven useful for
studying magnetic states in the Hubbard model within the
dynamical mean field theory \cite{zitzler.02}.

\begin{figure}[h]
\centerline{\includegraphics[width=6cm]{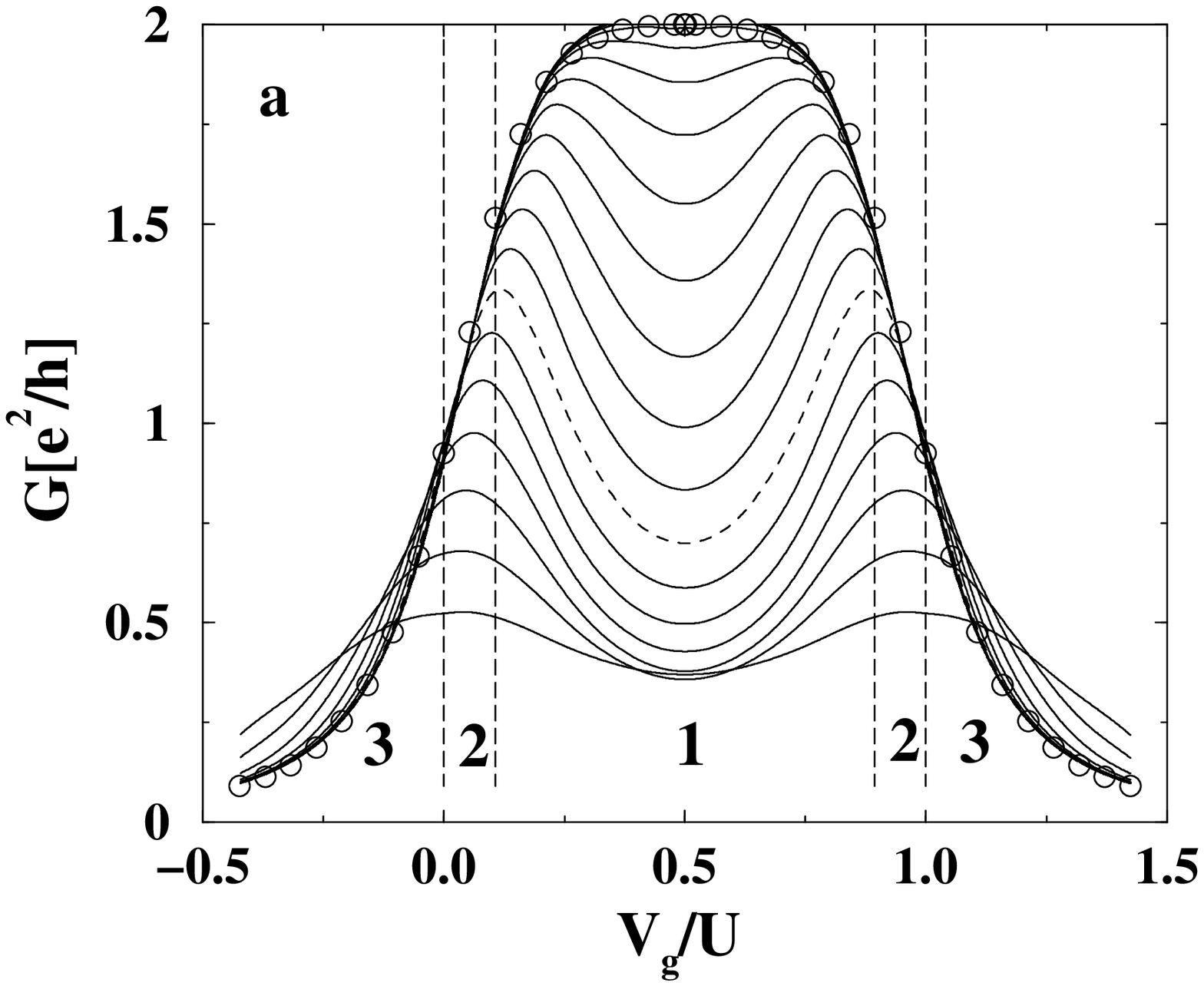}
\includegraphics[width=6cm]{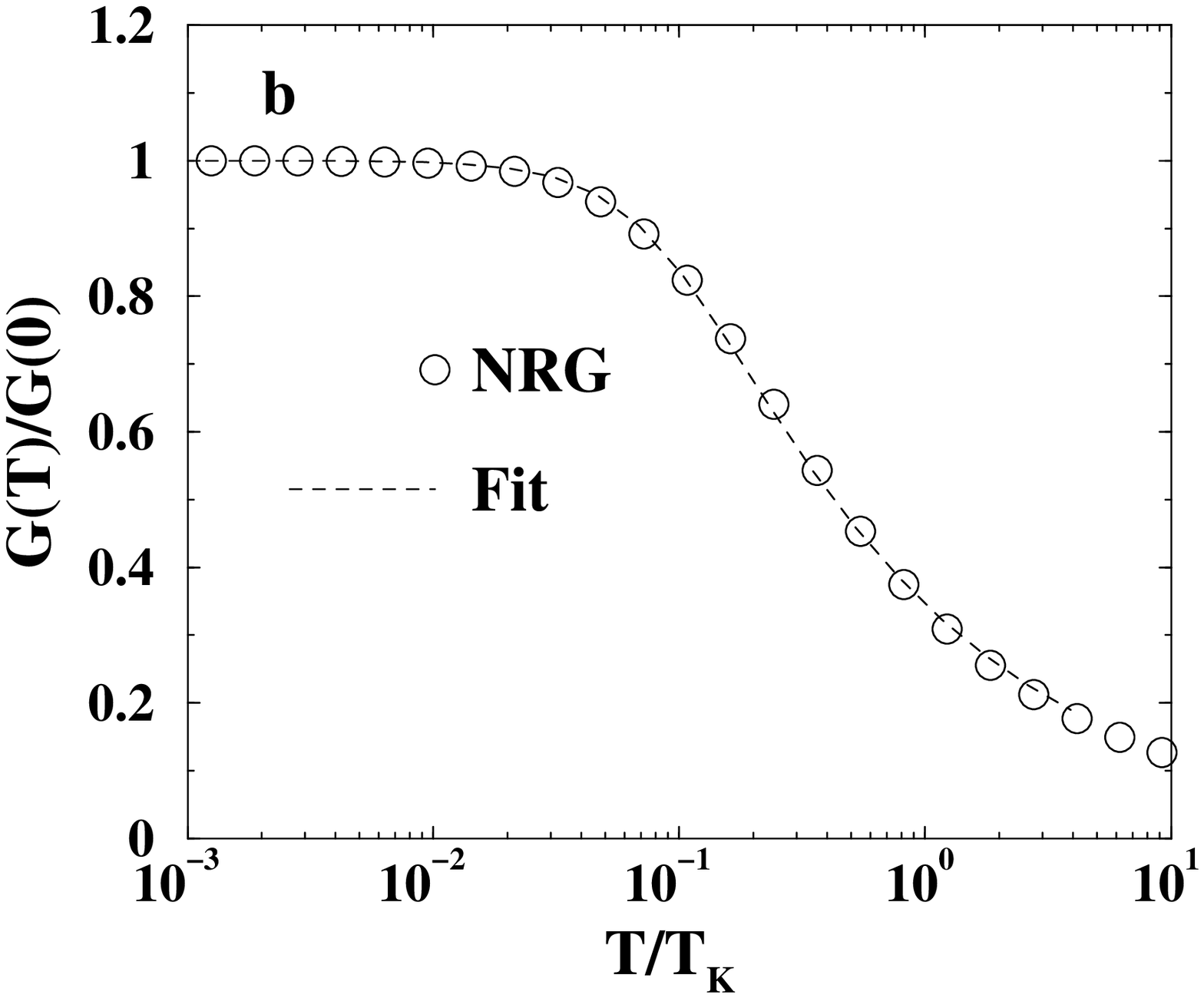}}
\caption{
(a) $G(T,H=0)$ versus $V_g$ in the regime, $T\leq \Gamma$, of strong quantum
fluctuations, for $U/\Gamma = 4.712$.
The symbols are the $T=0$ limit, Eq.\ (\ref{eq:unitarity}). 
Temperatures decrease from bottom to top and 
correspond to $T_N=0.64\Gamma\Lambda^{-N}, N=0,1,2,\dots,\Lambda=1.5$.
The dashed curve has $T=1.2T_{K}$, where $T_{K}\approx 0.05\Gamma$, 
is the HWHM of the $T=0$ Kondo resonance at mid-valley gate voltage.
The regions marked $1,2$ and $3$ and separated by vertical dashed lines 
correspond to the Kondo ($\varepsilon_d \lsim -0.5\Gamma$, i.e. $n_d \approx 1$), 
mixed valent ($|\varepsilon_d|\lsim 0.5\Gamma$, i.e. $n_d\approx 0.5$, or
by particle-hole symmetry, $|\varepsilon_d + U|\lsim 0.5\Gamma$, i.e. 
$n_d \approx 1.5$), and empty (full) orbital regimes 
($\varepsilon_d \gsim 0.5\Gamma$, i.e. 
$n_d\approx 0$, or, by particle-hole symmetry,
$\varepsilon_d + U\lsim -0.5\Gamma$, i.e. $n_d\approx 2$).
(b) The universal conductance curve $G(T)/G(0)$ for a quantum dot 
in the Kondo regime at mid-valley, $V_g/U=0.5$, (circles)
\cite{costi.94,costi.00}.
The dashed line is the fit formula  $G(T)/G(0) = (T_K'^2/(T^2
+T_K'^2))^s$ with $s=0.22$ and $T_K'=T_K/\sqrt{2^{1/s} -1}$
used in \cite{goldhaber.98} to 
interpolate the NRG results up to $10T_K$ (with
$T_K$ as in figure:~\ref{figure2}a).
}
\label{figure2}
\end{figure}
\section{Linear conductance for $H=0$}
\label{zero-field}
The gate voltage dependence of the linear
conductance of a quantum dot is shown in figure:~\ref{figure2}a for a decreasing set of 
temperatures. There are three distinct physical regimes, corresponding to three 
ranges of the gate voltage controlling $\varepsilon_d$ relative to the Fermi level
(see also figure:~\ref{figure3}a-c). 
In the empty (full) orbital regime (region 3) the conductance shows the expected
activated behaviour as a function of temperature. 
In the mixed valent regime (region 2), the behaviour of $G$ versus $T$ 
is approximately that corresponding to tunneling through a
resonant level close to the Fermi level. The most dramatic 
behaviour is in the Kondo regime
(region 1), where one sees an anomalous enhancement of the conductance with decreasing
temperature. The conductance continues to increase and eventually reaches the unitarity
limit of $2e^2/h$ (see below) at mid-valley ($V_g/U=0.5$ or $n_d=1$). This 
remarkable enhancement of the conductance results from the formation, with 
decreasing temperature, of the many-body Kondo resonance at the Fermi level of the leads.
The three types of behaviour have been observed in experiments on quantum dots 
carried out at $T\lsim \Gamma$ \cite{goldhaber.98,wiel.00,nygard.00}. 
The unitarity limit of the conductance, $2e^{2}/h$, has also recently been observed
\cite{wiel.00}. The $T\rightarrow 0$ conductance curve
depends only on $n_d$ via the Friedel sum rule \cite{glazman.88,ng.88,
langreth.66}
\begin{eqnarray}
G(T=0,H=0)&=&\frac{2e^2}{h}\sin^{2}(\pi n_d/2),\label{eq:unitarity}
\end{eqnarray}
and is shown in figure:~\ref{figure2}a (circles), with $n_{d}$ 
deduced from the spectral densities as a check on the calculations. 
\section{Scaling of the linear conductance at $H=0$}
\label{scaling}
In the Kondo regime, $G(T)/G(0)$, 
is a universal function of $T/T_K$ starting from low temperatures and 
extending up to some high temperature which depends on other details such as 
the precise value of $\Gamma$ and $\varepsilon_d$. The latter energy scales
cut off the universal behaviour of the conductance on the high temperature 
side. The universal conductance curves for the Kondo and Anderson 
models have been calculated for the experimentally interesting temperature 
range $0\leq T\leq 10 T_K$ via the NRG \cite{costi.94,costi.00,costi.01}. This is 
shown for the Kondo model in figure:~\ref{figure2}b together with an approximate 
interpolating formula used in \cite{goldhaber.98}. 
The scaling of $G(T)/G(0)$ with $T/T_K$ shown in figure:~\ref{figure2}b persists 
in the Anderson model throughout the Kondo regime (region 1) as long as
$T\ll \Gamma$. At higher temperatures and in the other regimes, the
conductance curves deviate from this universal shape (see Sec.~\ref{comparison} 
below). The Kondo scaling of the conductance in figure:~\ref{figure2}b
agrees well with measurements for both quantum dots in heterostructures
\cite{goldhaber.98,wiel.00} and and carbon nanotubes \cite{nygard.00}.  

\section{Comparison with experiment}
\label{comparison}
\begin{figure}[h]
\centerline{\includegraphics[width=6cm]{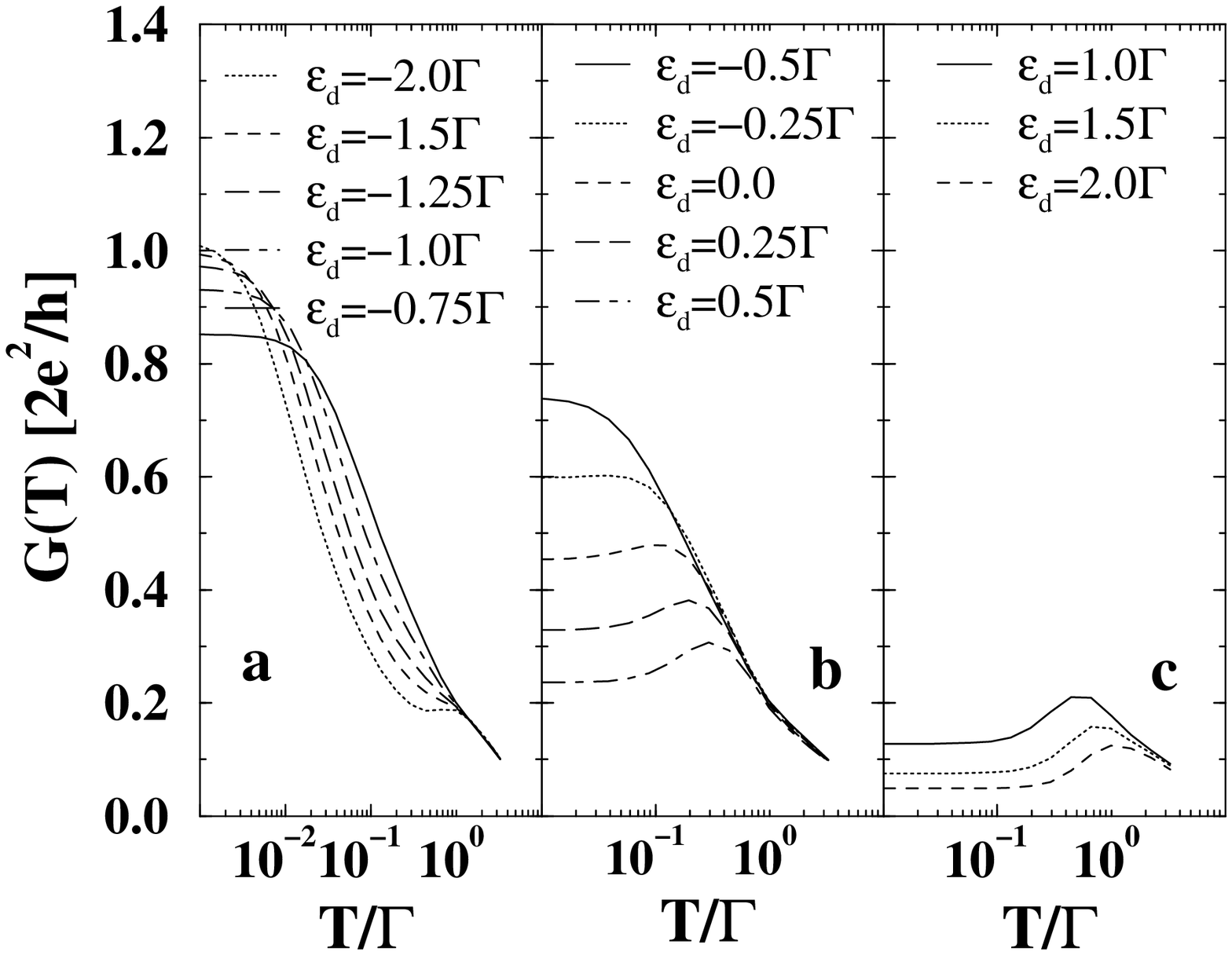}
\includegraphics[width=6cm]{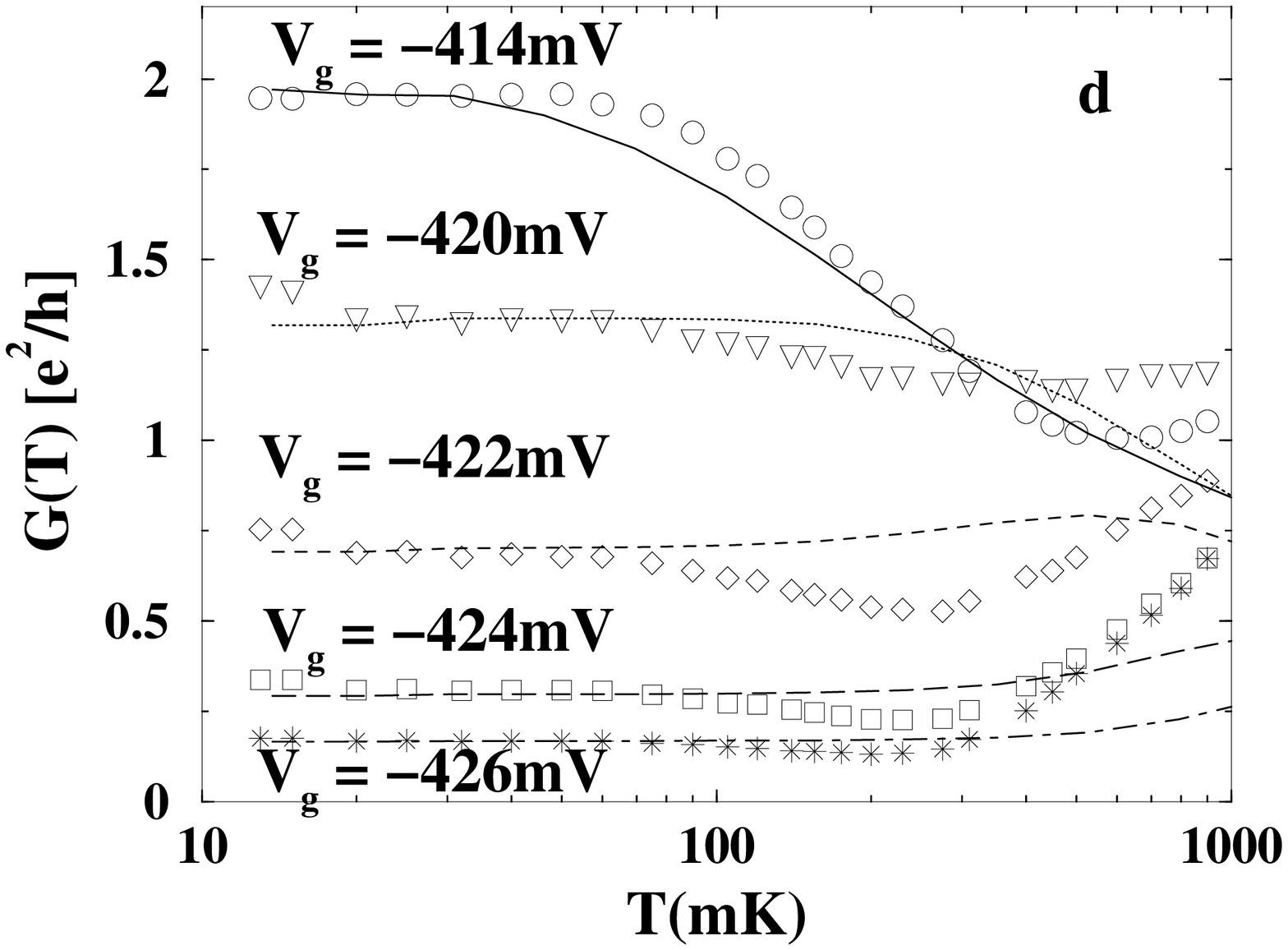}
}
\caption{ (a-c) $G(T)$ for gate voltages 
$V_g=-\varepsilon_d/e$ in, (a), the Kondo (K) regime, (b), the 
mixed valent (MV) regime ($\varepsilon_d=-0.5\Gamma,..,+0.5\Gamma$), and, 
(c), the empty orbital (EO) regime ($\varepsilon_d=+\Gamma,..,+2.0\Gamma$), 
and parameters as in figure:~\ref{figure2}a.
(d) Comparison of theoretical (lines) to experimental (symbols) 
conductance curves, $G(T)$, for the quantum dot in \cite{wiel.00} 
with $U=0.5$meV, $\Gamma=0.231$meV and $13$mK $\leq T \leq$ $900$mK. 
The values used for $\varepsilon_d$ in the calculations were
$\varepsilon_d=-0.85\Gamma$ ($V_g = -414$mV, K),
$\varepsilon_d=-0.25\Gamma$ ($V_g = -420$mV, MV),
$\varepsilon_d=+0.375\Gamma$ ($V_g = -422$mV, MV),
$\varepsilon_d=+1.00\Gamma$ ($V_g = -424$mV, EO),
$\varepsilon_d=+1.50\Gamma$ ($V_g = -426$mV, EO).
We find a linear dependence $V_g/e =-\alpha\varepsilon_d$ for
$V_g\leq -420$mV, which is a consistency check for determining 
$\varepsilon_d$ from $G(T\rightarrow 0)$ using Eq.~\ref{eq:unitarity}.
}
\label{figure3}
\end{figure}

For a comparison of theory with experiment in all regimes, the complete set
of conductance curves for the Anderson model is required. These are
shown in figure:~\ref{figure3}a-c. 
$G(T)$ in the empty orbital and mixed valent regimes has been calculated in
\cite{schoeller.00} and used to compare with experimental data in
\cite{goldhaber.98}. Here we make a parameter free comparison to 
similar data from Reference~\cite{wiel.00} for all three 
regimes of interest. The results are shown in figure:~\ref{figure3}d. 
In making the comparisons, we estimated $G(T=0)$ from $G(T=30mK)$, close
to the lowest effective electron temperature $T=40$mK \cite{wiel.00}, 
and used it to determine $n_d$ from  Eq:\ref{eq:unitarity} 
and hence the appropriate $\varepsilon_d$ to use in the NRG calculation for
$G(T)$ at all $T$. The calculations also used the values of
$\Gamma=231.4{\mu}eV$ and $U\approx 0.5$meV from the experiments.
It is remarkable that this zero parameter comparison yields the agreement
seen. The Kondo scale automatically comes out correctly as seen for the
$V_g=-414$mV comparison, and the agreement with the theoretical conductance
curve is very good up to $700$mK. The conductance curve used here 
includes the non-universal corrections discussed above and therefore
differs slightly from that used in \cite{wiel.00} (notably a slope change 
at $500$mK due to corrections from charge fluctuations). The general trends
of the experimental data in going to the mixed valent ($V_g=-420$mV, 
$V_g=-422$mV) and empty orbital cases ($V_g=-424$mV, $V_g=-426$mV)
are well reproduced by the calculations. In particular the expected
finite $T$ peak in $G(T)$ \cite{schoeller.00} develops and becomes 
increasingly more pronounced on
entering the  empty orbital regime. As described earlier, transport in this
regime is likely to involve additional neighbouring levels and a larger
conductance at higher temperatures, as observed in the experiments.
In this light, the main discrepancy remaining between theory and experiment 
appears to be the dip in $G(T)$ at $200$-$300$mK in the measurements. No 
signature of this is present in our model calculations. This could be due to
interference effects associated with more than one level. 
It would therefore be interesting to investigate this possibility further.

\section{Effect of a magnetic field}
\label{magnetic-field-effects}
\begin{figure}[h]
\centerline{\includegraphics[width=6cm]{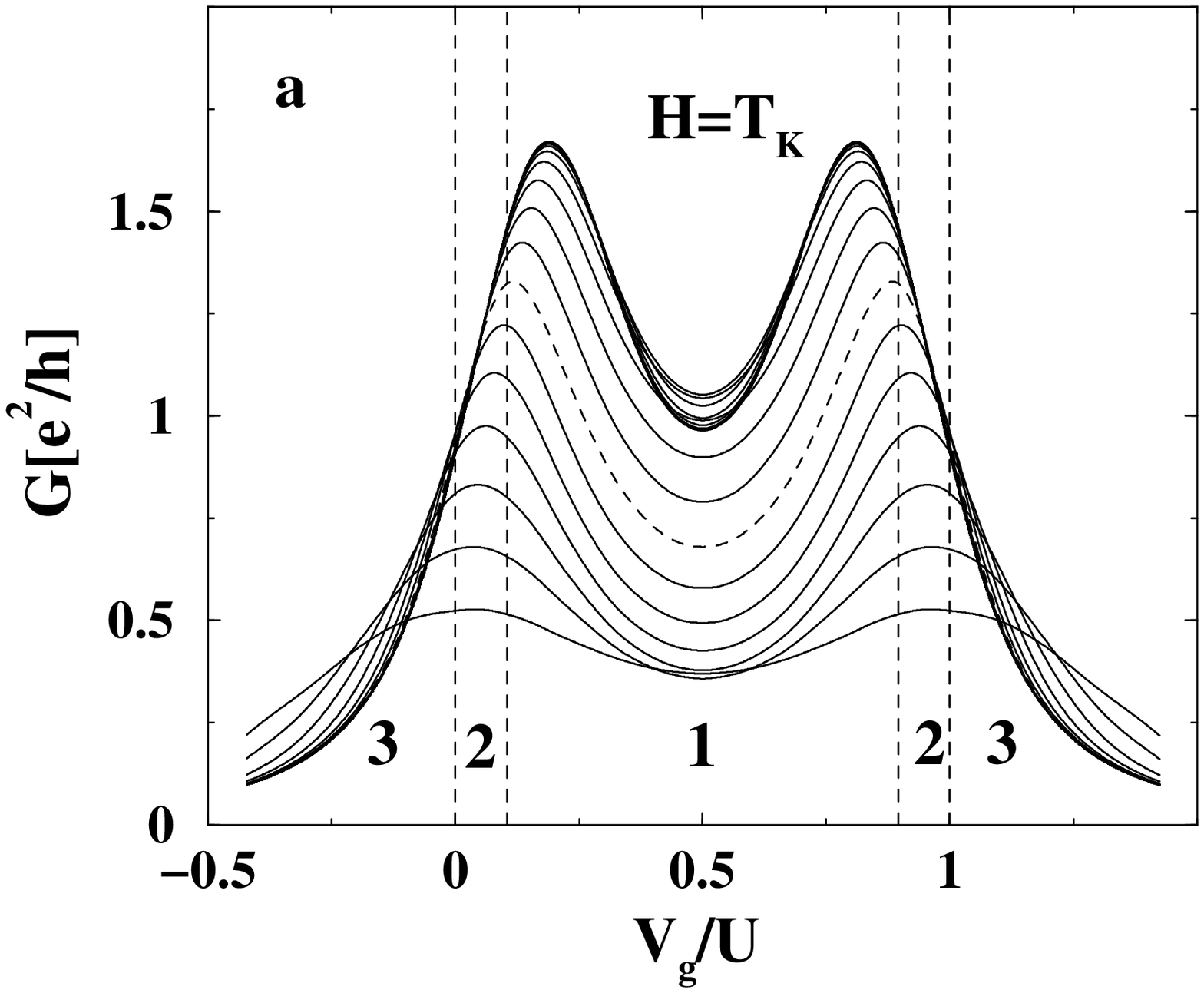}
\includegraphics[width=6cm]{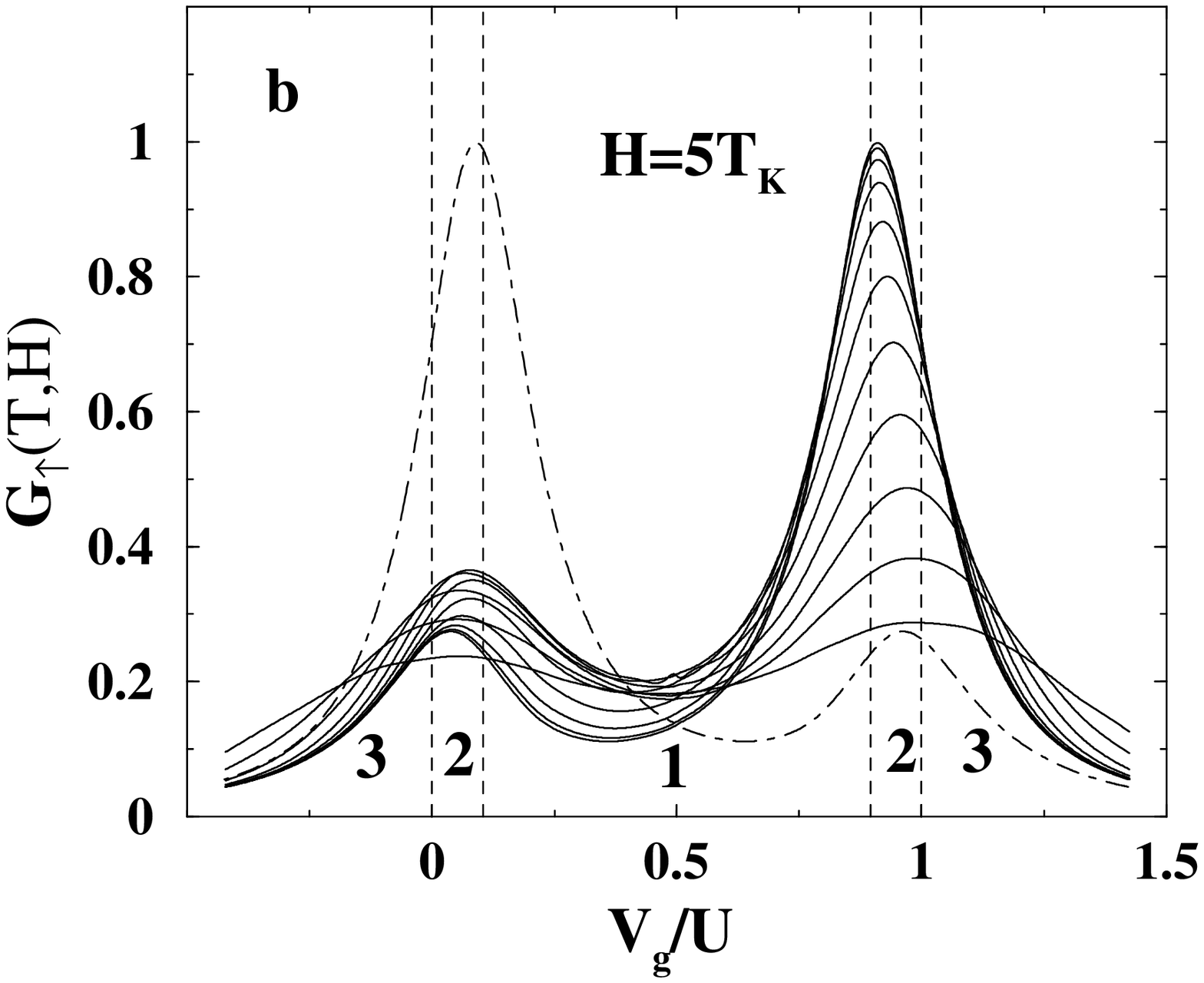}
}
\caption{
(a) $T$ and $V_g$ dependence of $G(T,H)$ 
for $H=T_K$ and for temperatures and parameters as in figure:\ref{figure2}a.
(b) $T$ and $V_g$ dependence of 
the spin-resolved conductance $G_{\uparrow}(T,H)$ 
for $H=5T_K$ and for temperatures and parameters
as in figure:\ref{figure2}a. The dot-dashed curve is for
the down spin conductance, $G_{\downarrow}(T,H)$, at the
lowest temperature.
}
\label{figure4}
\end{figure}

A magnetic field suppresses the Kondo effect on a scale of $T_K$
\cite{hewson.93} and consequently it is expected to have a drastic effect
on the conductance of a quantum dot \cite{costi.00,costi.01}.
Indeed, figure:~\ref{figure4}a shows that a field of $H=T_K$ suffices to reduce the
$T=0$ conductance at mid-valley to nearly half its value at $H=0$ (cf. 
figure:~\ref{figure2}a). This can be understood from the splitting of the
Kondo resonance in a magnetic field and its strong reduction at the Fermi level
(see figure:~\ref{figure1}b and \cite{costi.00}). The suppression of the conductance
is large in the whole Kondo regime (region 1) as long as $T\lsim T_K$. At higher
temperatures and in the other regimes, the effect of a field on the total
conductance is less pronounced. However, in these regimes too, there can be 
a large effect in the {\em spin-resolved} conductances (see below).

The effects of a magnetic field become even more 
apparent in spin-resolved quantities such as $G_{\uparrow}$ in 
figure:~\ref{figure4}b
(by particle-hole symmetry $G_{\downarrow}$ is the mirror reflection 
of $G_{\uparrow}$ about $V_g/U=0.5$). 
As in $G$, a magnetic field has a large effect
on the spin-resolved conductance in the Kondo regime. In addition, there
is quite a dramatic spin-filtering effect in the mixed valent regime,
with $|G_{\uparrow}(0,H)-G_{\downarrow}(0,H)|$ being largest in this
regime. In contrast, both up and down spin conductances are approximately
equally suppressed in the other regimes. A quantum dot in a field is
seen to act as a spin-filter as discussed in \cite{recher.00} for
dots very weakly coupled to leads ($G_{\sigma}\ll e^2/h$). 

The field dependence of the conductance of quantum dots defined in 
carbon nanotubes has been studied \cite{nygard.00} and a 
preliminary comparison between our results and the experimental 
$G(B,T\lsim T_K)$ shows good agreement \cite{nygard.02}.

\section{Conclusions} 

We considered electronic transport through a strongly
interacting quantum dot in zero and finite magnetic fields and in the
low temperature regime $T\lsim \Gamma$ where quantum fluctuations strongly
modify the transport for an odd number of electrons on the dot. 

The NRG conductance curves, $G(T,V_g)$, were used
to make a parameter free comparison with experimental results of Reference 
\cite{wiel.00} and we found good agreement. Discrepancies in the empty 
orbital regime where attributed to the neglect, within our model, of
neighbouring levels

The results for the magnetoconductance in the Kondo regime exhibited 
the strong suppression of the Kondo effect by a magnetic field. A
large spin-filtering effect was found in the mixed valent regime.

The field of non-equilibrium transport through quantum dots remains largely
unexplored. Perturbative methods valid in the limit of large transport voltage, 
and magnetic field, $V,H\gg T_K$, are being developed \cite{rosch.02}, but 
non-perturbative techniques, such as extensions of NRG, will be required 
to address strong coupling.

\begin{acknowledgements}
I would like to thank the authors of Reference~\cite{wiel.00} for sending me 
the experimental data used in figure:~\ref{figure3}b and W. van der Wiel for
useful discussions. Financial support from the Deutsche 
Forschungsgemeinschaft, 
and in part by the National Science Foundation under Grant No. PHY99-07949 
during the writing of this paper at the KITP of UCSB, is acknowledged. 
\end{acknowledgements}

\end{article}

\end{document}